\documentstyle[preprint,tighten,aps,psfig]{revtex}
\begin{document}           %
\draft
\preprint{\vbox{\noindent
To appear in Astroparticle Physics \hfill hep-ph/9904225\\
          \null\hfill  INFNCA-TH9903}}
\title{Vacuum oscillations and excess of high energy solar neutrino 
events observed in Superkamiokande
  }
\author{V. Berezinsky$^{1,}$\cite{email1},
         G.~Fiorentini$^{2,}$\cite{email2},
         and M.~Lissia$^{3,}$\cite{email3}}
\address{
$^{1}$ Istituto Nazionale di Fisica Nucleare, Laboratori Nazionali del
       Gran Sasso, I-67010 Assergi (AQ),Italy and Institute for Nuclear 
Research of RAN, Moscow, Russia\\
$^{2}$ Dipartimento di Fisica dell'Universit\`a di Ferrara and Istituto
      Nazionale di Fisica Nucleare, Sezione di Ferrara, 
      via Paradiso 12, I-44100 Ferrara, Italy\\
$^{3}$ Istituto Nazionale di Fisica Nucleare, Sezione di Cagliari and
      Dipartimento di Fisica dell'Universit\`a di Cagliari,
      I-09042 Monserrato (CA), Italy}
%
\date{3 April 1999; references and notes added 5 August 1999}
\maketitle                 
\begin{abstract}
The excess of solar-neutrino events above 13~MeV that has been recently
observed by Superkamiokande can be explained by the vacuum oscillation 
solution to the Solar Neutrino Problem (SNP).
If the boron neutrino flux is 20\% smaller than the standard solar
model (SSM) prediction and the chlorine signal is assumed 30\% 
(or $3.4 \sigma$) higher
than the measured one, there exists a vacuum oscillation  solution to SNP that 
reproduces both
the observed spectrum of the recoil electrons, including the high energy 
distortion, and the other measured neutrino rates.
The  most distinct signature of this solution is a semi-annual seasonal
variation of the $^7$Be neutrino flux with maximal amplitude. While
the temporal series of the GALLEX and Homestake signals suggest that
such a seasonal variation could be present, future detectors (BOREXINO, LENS
and probably GNO) will be able to test it.
\end{abstract}
%
%
\narrowtext
\section*{}
Superkamiokande~\cite{REF:SKAM,REF:Totsuka,REF:Inou} has recently observed
an excess of solar-neutrino events at electron energies higher than $13~$MeV.
This excess cannot be interpreted as a distortion of the boron
neutrino spectrum due to neutrino 
oscillations~\cite{REF:SKAM,REF:Totsuka,REF:Inou,REF:BKS},
if one restricts oneself to those oscillation solutions that
explain the observed gallium, chlorine and water-cerenkov neutrino rates.

It is tempting to think that this excess is the result of low statistics 
or small systematic errors 
at the end of the boron neutrino spectrum. For example, 
because of very steep end of the electron spectrum, even  small 
systematic error in electron energy ({\em e.g.}, due to calibration)
could enhance the number of events in the highest energy bins.
One should wait for future Superkamiokande data, where such possible
systematic effects will be further elaborated. The data from 
the SNO detector, which will come in the operation soon, {\em e.g.},
see Ref.~\cite{REF:SNO}, can shed light on this excess.

Another possible explanation of this excess~\cite{escri,REF:BaKr} is that the 
Hep neutrino flux might be significantly larger (about a factor 10--20) 
than the
SSM prediction. The Hep flux depends on solar properties, such as the
$^3$He abundance and  the temperature, and on $S_{13}$, the zero-energy 
astrophysical $S$-factor of the 
\mbox{$p+{}^3He \rightarrow {}^4He + e^+ + \nu$}
reaction. Both SSM based~\cite{REF:BaKr} and model-independent~\cite{REF:BDFR}
approaches give a robust prediction for the ratio $\Phi_{\nu}(Hep)/S_{13}$.
Therefore, this scenario implies a cross-section larger by
a factor 10--20  than the present calculations (for
reviews see~\cite{REF:BaKr,INT}). Such a large correction to the calculation
does not seem likely, though it is not excluded. 
A large Hep neutrino flux remains a possible
explanation of the excess. The signature of Hep neutrinos, the presence of
electrons above the maximum boron neutrino energy, can be tested by
the SNO experiment.

The Superkamiokande collaboration noticed~\cite{REF:SKAM} that vacuum
oscillations with large $\Delta m^2$ explain the observed high-energy 
excess. However, the same Ref.~\cite{REF:SKAM} emphasizes that those
oscillation parameters that reproduce the excess do not solve the Solar
Neutrino Problem (SNP), {\em i.e.}, they do not
explain the global rates observed by the four solar neutrino experiments. In 
Ref.~\cite{REF:dvo} it was demonstrated that if the SSM prediction for 
the boron neutrino flux is reduced by factor $f_B= 0.8$ and the chlorine
experimental signal is arbitrarily assumed to be larger by a factor
$f_{Cl} =1.3$, the vacuum oscillation solution to SNP (global rates) 
corresponds to  
$\Delta m^2 = 4.2\cdot 10^{-10}$~eV$^2$ and $\sin^2 2\theta=0.93$: this
choice of parameters reproduces also the excess of high-energy
recoil electrons in the Superkamiokande spectrum~\cite{REF:SKAM}.
In this paper we shall further elaborate upon this specific vacuum
oscillation solution. For the sake of conciseness, we shall refer to this
solution ($\Delta m^2 = 4.2\cdot 10^{-10}$~eV$^2$ and $\sin^2 2\theta=0.93$)
as HEE (High-Energy Excess) VO.

We shall start with a short description of relevant features of those
vacuum oscillation solutions, whose parameters fit all global rates; these
solutions will be indicated altogether as VO.
 
Vacuum oscillations can reconcile the SSM with the observed rates of 
all three kinds of solar neutrino experiments (for reviews
see~\cite{REF:Bahc,REF:BiPe,REF:Turck,REF:Hax}). A recent detailed
study~\cite{REF:HaLa95,REF:BK96,REF:KP96,REF:HaLa97,REF:Lisi} of VO solutions 
shows that global fits to the data
result in oscillation parameters 
within the ranges 
$5\cdot 10^{-11}$~eV$^2\leq \Delta m^2 \leq 1\cdot 10^{-10}$~eV$^2$ 
and $0.7\leq\sin^2 2\theta\leq 1$ for oscillations between active 
neutrinos.
The large range of $\sin^2 2\theta$ is mainly caused by uncertainties in the 
B-neutrino flux, though other uncertainties contribute too;
$\Delta m^2$ is much less sensitive to changes of the B-neutrino flux. 
These  effects have been explicitly investigated in 
Refs.~\cite{REF:KrSm,REF:KP96}. In the SSM the B-neutrino flux uncertainties
($+19\%,-14\%, $~\cite{REF:BP98}) are mainly caused by the uncertainties
in  $S_{17}$ (the $p$-Be cross section is poorly known)
and by the strong temperature dependence of this flux.
The above uncertainties are only $1\sigma$ errors and the actual
discrepancy could be larger, especially due to
the $S_{17}$ factor. This large uncertainty of the B-neutrino flux has
motivated  several authors to consider the boron flux as
$\Phi_B=f_B\Phi_B^{SSM}$ with $f_B$ as a free
parameter~\cite{REF:BKS,REF:KrSm,REF:HaLa95,REF:KP96}. 

A signature of vacuum oscillation is the anomalous seasonal variation 
of the neutrino flux at low energies~\cite{REF:Pom,REF:BiPe}. 
The distance between the Sun and the Earth varies during the year by
about 3\% affecting the detected flux both because of the $1/r^2$ geometrical
factor and because of the dependence of the survival probability
$P(\nu_e \to \nu_e)$ on the distance. The second effect is
absent for MSW solutions. 
Nevertheless, the MSW solution also
predicts seasonal variations of neutrino flux, which are connected with 
the day/night effect and are caused by the longer winter nights
(for recent calculations see~\cite{Valle,BKS1} and references to
earlier works therein). The MSW seasonal variations are weaker than
the VO ones at low energies. In the case of VO, the monochromatic
Be-neutrinos are expected to show the strongest seasonal
variations~\cite{REF:BiPo,REF:GK,REF:KrPe95,REF:Lisi}; on the contrary,
Be-neutrinos should show very small seasonal variations in the case of
MSW oscillations. Since Be-neutrinos  are monochromatic, their flux
shows the entire seasonal variation predicted by VO;
the effect is reduced for the other fluxes due to the averaging over
the different phases of neutrinos with different energies within the
interval of observation, $\Delta E$.

Seasonal variations 
for $\Delta m^2$ larger than the values allowed by VO solutions were 
recently analyzed in Ref.~\cite{REF:GKK}. The authors found some significant 
consequences such as energy dependence and correlation with distortion of 
the spectrum. The latter effect was also discussed earlier in
Ref.~\cite{REF:MS}. In relation to the chlorine signal, seasonal variations
were analysed in early work~\cite{REF:EHRLICH}. A clear discussion
of the seasonal variation effect has been presented in
Refs.~\cite{REF:Rosen}.

To explain the excess in electron spectrum observed by Superkamiokande
we allow a boron neutrino flux 15--20\% smaller than the SSM 
prediction, and we allow that the chlorine signal be about 30\%
larger than the Homestake observation.
This assumed $3.4\sigma$ increase could have a combined statistical 
and systematic origin  though we do not have any concrete argument in favor of
such systematic error in the Homestake experiment.

In our calculation, we shall use neutrino fluxes from the BP98
model~\cite{REF:BP98} with the B-neutrino flux rescaled as 
$\Phi_B= f_B \Phi_B^{SSM}$. 

For the chlorine rate we assume $R_{Cl} = 2.56 f_{Cl}$~SNU (the Homestake 
experiment gives the rate~\cite{REF:Hom}  $2.56 \pm 0.16 \pm 0.16$~SNU).
It is easy 
to see that for $f_{Cl}=1.3$ the assumed signal 3.33~SNU is $3.4\sigma$ 
higher than one given by Homestake (systematic and statistical error
are incoherently combined). 

For the gallium
rate we use the average of the GALLEX~\cite{REF:Ki} and SAGE~\cite{REF:Ga} 
results:
$72.5 \pm 5.7$~SNU. Finally, we take the Superkamiokande
result~\cite{REF:SKAM}:
$(2.46 \pm 0.09)\cdot 10^6$~cm$^{-2}$s$^{-1}$.
For each pair $f_B$ and $f_{Cl}$
we find the VO solution, {\em i.e.}, the parameters
($\Delta m^2, \sin^22\theta$), that explain
the observed rates, and then we calculate the corresponding 
boron neutrino spectrum.

For example, for $f_B=0.8$ and $f_{Cl}=1.3$ the oscillation 
parameters ($\Delta m^2=4.2 \cdot 10^{-10}$~eV$^2, \sin^22\theta=0.93$)
give a good fit to all rates ($\chi^2$/d.o.f. = 3.0/3): This is not the best
fit point, which has $\chi^2\approx 0$, therefore the 3 d.o.f. are the three
experimental rates. On the other hand,
the spectrum with these oscillation parameters reproduces~\cite{REF:SKAM} 
the excess of high-energy events observed in the Superkamiokande spectrum.
More generally,
this choice of oscillation parameters gives rates in agreement with the
experiments at the $2\sigma$ level for $0.77 \leq f_B \leq 0.83$ and
$1.3 \leq f_{Cl} \leq 1.55$.

In Fig.~1 we present the neutrino-induced electron spectra for the vacuum 
oscillation solutions as the ratio to the
SSM unmodified spectrum~\cite{REF:BP98}. The dotted and dashed curves show
two spectra corresponding to the VO solutions of Ref.~\cite{REF:BKS} and
Ref.~\cite{REF:HaLa97}, respectively. The solid line shows the VO 
oscillation solution that is discussed in this paper (HEE VO) corresponding to
$\Delta m^2=4.2\cdot 10^{-10}$~eV$^2$ and $\sin^2 2\theta=0.93$  
($f_B=0.8$ and $f_{Cl}=1.3$).

 The role of the two parameters, $f_B$ and $f_{Cl}$, for the best fit 
of the spectrum is
different: while $f_B$ mostly changes $\sin^2 2\theta$, $f_{Cl}$
affects $\Delta m^2$ and, therefore, the spectrum. Values of $f_{Cl}$
as low as 1.2 already give a bad fit to the observed spectrum.   

The anomalous seasonal variations of Be-neutrino flux and of the gallium
signal are shown in the Fig.~2 (see also \cite{REF:dvo}). Anomalous seasonal 
variation is described by the survival probability of the electron neutrino 
$P(\nu_e \to \nu_e)$. For Be-neutrinos with energy $E=0.862$~MeV 
the survival probability (the suppression factor for electron neutrinos) is 
given by
\begin{equation}
\label{EQ1}
P(\nu_e \to \nu_e)= 1-\sin^2 2\theta
\sin^2\left( \frac{\Delta m^2 a}{4E}\, (1+e\cos \frac{2\pi t}{T} ) \, \right)
\, ,
\end{equation}
where $a=1.496\cdot 10^{13}$~cm is the semimajor axis, 
$e=0.01675$ is the eccentricity of the Earth's orbit, and $T=1$~yr is the 
orbital period. The phase in 
Eq.~(\ref{EQ1}) is such that $t=0$ corresponds to the aphelion.
In Fig.~2 the solid and dashed curves show the variation of the Be-neutrino
flux for the HEE VO and VO~\cite{REF:BKS} cases, respectively. The case of
the HEE VO (solid curve) is dramatically different from the VO case: there are 
two maxima and minima during one year and the survival probability oscillates 
between 
$1-\sin^2 2\theta \approx 0.14$ and 1. The explanation is obvious: the HEE VO 
solution has a large $\Delta m^2$, which results in a phase
$\Delta m^2 a/(4E) \approx 93$, large enough to produce two full harmonics 
during one year, when the phase changes by about 3\% due to the factor
$(1+e\cos 2\pi t/T)$. 
The flat central maximum with a shallow local minimum has
a trivial origin: the extrema of $P(\nu_e \to \nu_e)$ in Eq.~(\ref{EQ1}) 
correspond to phases $k\pi/2$, where k are integers, and to the phases with 
$\cos 2\pi t/T = \pm 1$. The accidental proximity of these phases can result 
in three nearby extrema. The shallow minimum in Fig.~2 disappears with small 
changes in $\Delta m^2$.

The phases of maxima and minima in terms of t/T 
are not fixed in the HEE VO solution, because tiny changes of
$\Delta m^2$ shift their positions: {\em e.g.}, 1\% change in $\Delta m^2$ 
shifts the position of an extremum by more than one month
(see Eq.~(\ref{EQ1})).
 
As one can see from Fig.~2, the HEE VO solution predicts that the beryllium 
electron neutrinos should arrive almost unsuppressed during about four months 
in a year!

According to the SSM, beryllium neutrinos contribute 34.4~SNU out of
the total gallium signal of 129~SNU.
Therefore, the strong $^7$Be neutrino oscillation
predicted by the HEE VO solution also implies an appreciable variation of 
total gallium signal. In Fig.~2 the dotted curve shows this variation
corresponding to the HEE VO solution, which can be compared with the
weaker variation corresponding to the best-fit VO solution (dashed-dotted 
curve).
It is possible that the HEE VO variation could already be partially testable
by the existing gallium data, and this possibility will significantly
increase when the results from GNO with its larger statistics are available.

In Fig.~3 the predicted time variation of the gallium signal is compared 
with GALLEX data (see also \cite{REF:dvo}). GALLEX data have been
analysed according to the time of
the year of the exposures and grouped in six two-month bins
(M.~Cribier cited in \cite{REF:Ki}): the data points with error bars
in Fig.~3 reproduce the result of this analysis. 
The data give the 
rates averaged for the same two months every year of observations. 
The theoretical prediction (solid curve) is plotted with the same averaging. 
The $7\%$ geometrical variation is included. Both the 
phase of the time-variation and the average flux have been taken to fit 
the data. The fit by the theoretical curve has $\chi^2$/d.o.f.= 0.85/4; the 
fit by a nonoscillating signal is also good: $\chi^2$/d.o.f.=1.36/5. Because 
of the limited statistics, we 
do not interpret the good visual agreement  in Fig.~3 as a proof of HEE VO 
solution, though it is certainly suggestive.

The comparison of the predicted time variation with preliminary 
data \cite{REF:Ga} of the other 
gallium detector SAGE is shown in Fig.~4. Note that this time we can not 
choose 
the phase arbitrary: it is already fixed by the fit to the GALLEX data.
Because of the larger fluctuations of the SAGE data (compare Fig.~3 and
Fig.~4) the agreement with the predicted variation is worse.

In Fig.~5 the predicted variation is compared with the Homestake data 
(see Ref.~\cite{REF:EHRLICH} for an earlier analysis of indications for
seasonal oscillations in the Homestake data). The phase of the theoretical 
dependence is kept fixed at the value fitted to the GALLEX data. As for
the GALLEX data we find that the HEE VO theoretical curve gives a better
fit ($\chi^2$/d.o.f.=1.4/5) than the time-independent signal, which
however cannot be excluded ($\chi^2$/d.o.f.=3.1/5). This agreement is further 
strengthened by the fact that the phase of time dependence was not chosen 
to fit the Homestake data, since it was already fixed by the GALLEX data. One 
should consider this agreement
as additional indication for the HEE VO solution.

Finally, in Fig.~6 (Fig.~7) we compare the time variation of the
Superkamiokande signal for recoil electrons with energies higher than 10~MeV
(11.5~MeV) with the HEE VO predictions. The fit of the data is good: 
$\chi^2$/d.o.f.=2.7/7 ($\chi^2$/d.o.f.=5.1/7). Similar calculations were
done by the Superkamiokande collaboration~\cite{REF:Suzuki}, by A.~Smirnov
(private communication) and by M.~Maris and S.~Petcov~\cite{REF:Petcov}. 

While the agreement between the HEE VO solution and each single observational
datum on seasonal variations might appear accidental and not statistically
significant, the combined agreement with all data on seasonal
variations, as shown in Figs.~3--7 (total $\chi^2/d.o.f = 34.7/41$),
appears to be quite a suggestion in favour of the HEE VO solution.

As in our previous work~\cite{REF:dvo}, we prefer not to make a global fit 
in terms of $\chi^2/d.o.f.$ to all available data (rates, spectrum and time
variations). The large number of degrees of freedom can hide a discrepancy
with some particular data, especially if it corresponds to only one degree 
of freedom, like the chlorine rate in our case. A small $\chi^2$ is only
a necessary condition for the correct model. One can find such a global fit
in the paper by Barger and Whisnant~\cite{BarWhi}, which appeared after
this work was completed. The authors study the VO solution with  
$\Delta m^2=4.42\cdot 10^{-10}$~eV$^2$ and $\sin^2 2\theta=0.93$, {\em i.e.},
parameters close the ones we consider. They find this solution as
the global best fit to the rates, spectrum and time dependence of
SuperKamiokande signal ($\chi^2=39$ for 26 degrees of freedom).

In conclusion, the combination of a B-neutrino flux 20\% lower than in
the SSM (easily allowed by the present uncertainties) and of the assumption
that the chlorine signal be $30\%$  ($3.4\sigma$) higher than the one
presently observed by Homestake
results in a vacuum oscillation solution (HEE VO) that fits the electron
spectrum
recently observed by Superkamiokande. This solution predicts strong
seasonal variation of $^7$Be-neutrino flux: some indication to such a
variation is already seen in the GALLEX and Homestake
data. Seasonal dependence of the Superkamiokande data for
electron energies higher than 10~MeV and 11.5~MeV provide further 
indication in favour of the HEE VO solution.
The anomalous seasonal variation of Be-neutrino flux predicted by the HEE VO 
solution can be reliably observed by the future
BOREXINO~\cite{REF:BOREXINO} and LENS~\cite{REF:LENS} detectors. 
Additionally, LENS, which should measure the flux and spectrum of $pp$ 
neutrinos, will be able to observe the suppression of $pp$
neutrino flux, $P(\nu_e \to \nu_e)= 1- (1/2)\sin^2 2\theta=0.53$,
which is another signature of VO solutions.

\section*{Acknowledgements}
We are grateful to A.~Bettini, T.Kirsten and A.~Yu.~Smirnov for useful 
discussions. The participation of F.~Villante at some stage of work is 
gratefully acknowledged.

\begin{figure}
\caption[a]{
Ratio of the vacuum oscillation spectra to the SSM spectrum.
The solid curve corresponds to the HEE VO solution with
$\Delta m^2=4.2\cdot 10^{-10}$~eV$^2$ and $\sin^2 2\theta=0.93$.
The dashed and dotted curves correspond to the VO solutions
of Refs.~\cite{REF:HaLa97} and \cite{REF:BKS}, respectively.
Energy resolution is taken into account.
The data points show the 708-day Superkamiokande 
result~\cite{REF:Totsuka,REF:Inou}.
}
\end{figure}

\begin{figure}
\caption[b]{
Anomalous seasonal variations of the beryllium neutrino flux and gallium
signal for the VO and HEE VO solutions.
The survival probability $P(\nu_e\to \nu_e)$ for Be neutrinos  
is given for the HEE VO (solid curve) and the VO (dashed curve) 
solutions as function of time ($T$ is an orbital period). The dotted
(dash-dotted) curve shows the time variation of gallium signal in SNU for the
VO~\cite{REF:BKS} solutions.
           }
\end{figure}
\begin{figure}
\caption[c]{
Seasonal variations predicted by the HEE VO solution are compared with 
the GALLEX data. Theoretical dependence includes oscillations and $7\%$ 
geometrical effect. The global phase of oscillation (undefined in HEE VO) and 
the mean rate (taken in HEE VO as the average of the GALLEX and SAGE rates)
have been chosen to fit the data. The fit with the HEE VO solution has
$\chi^2$/d.o.f.=0.87/4, while the no-oscillation fit has
$\chi^2$/d.o.f.=1.36/5.
           }
\end{figure}

\begin{figure}
\caption[d]{
Seasonal variations predicted by the HEE VO solution are compared
with the SAGE preliminary data~\cite{REF:Ga}. Theoretical dependence 
includes oscillations 
and the $7\%$ geometrical variations. The phase of oscillation has
already been fixed by the GALLEX data, while the mean rate has been chosen
to fit the data. The fit with the HEE VO solution has
$\chi^2$/d.o.f.=8.9/5, while a time-independent fit gives
$\chi^2$/d.o.f.=3.8/5.
           }
\end{figure}

\begin{figure}
\caption[e]{
Seasonal variations predicted by the HEE VO solution are compared
with the Homestake data~\cite{REF:Hom} binned according to the mean
exposure time.
Data~\cite{REF:Hom} take into account the 7\% geometrical variation.
The phase of the oscillation has already been fixed to fit the 
GALLEX data, while the mean rate has been chosen to 
fit the data.   The fit with the HEE VO solution gives
$\chi^2$/d.o.f.=1.4/5, while a constant fit (no oscillation)
gives $\chi^2$/d.o.f.=3.1/5.
           }
\end{figure}

\begin{figure}
\caption[f]{
Seasonal variations predicted by the HEE VO solution (solid line) and
by the geometrical effect only (dashed line) are compared with 
the Superkamiokande data for electron recoil-energies $E_e  > 10$~MeV. 
The fit with the HEE VO solution gives
$\chi^2$/d.o.f. = 2.7/7, while the one with the geometrical effect only 
(no oscillation) gives $\chi^2$/d.o.f. = 2.3/7.
           }
\end{figure}

\begin{figure}
\caption[g]{
Seasonal variations predicted by the HEE VO solution (solid line) and
by the geometrical effect only (dashed line) are compared with 
the Superkamiokande data for electron recoil-energies $E_e  > 11.5$~MeV. 
The fit with the HEE VO solution gives
$\chi^2$/d.o.f. = 5.1/7, while the one with the geometrical effect only
(no oscillation) gives $\chi^2$/d.o.f. = 6.8/7.
           }
\end{figure}

\begin{figure}[c]
\psfig{figure=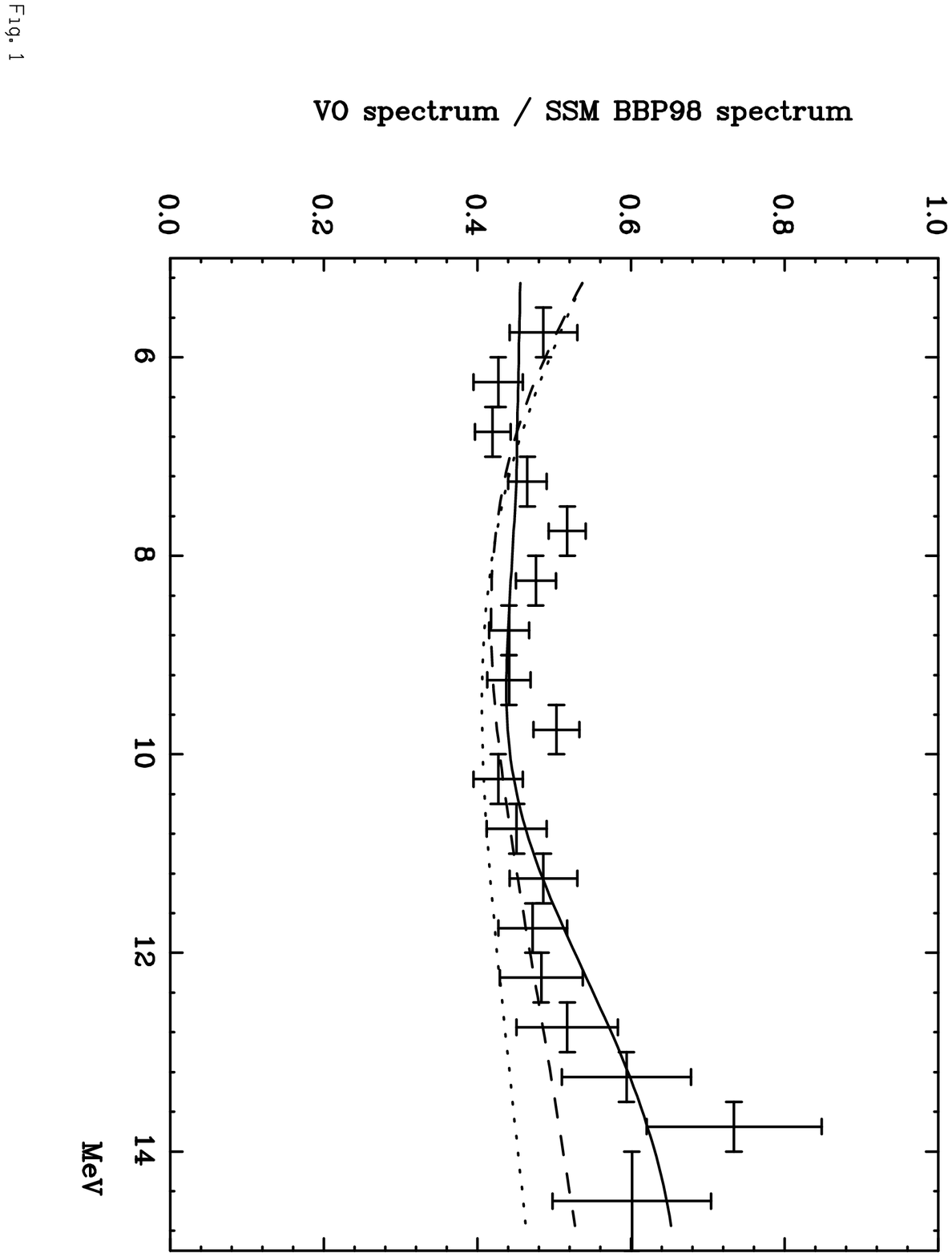,bbllx=36pt,bblly=36pt,bburx=576pt,bbury=756pt,%
height=22.5cm}
\end{figure}

\begin{figure}[c]
\psfig{figure=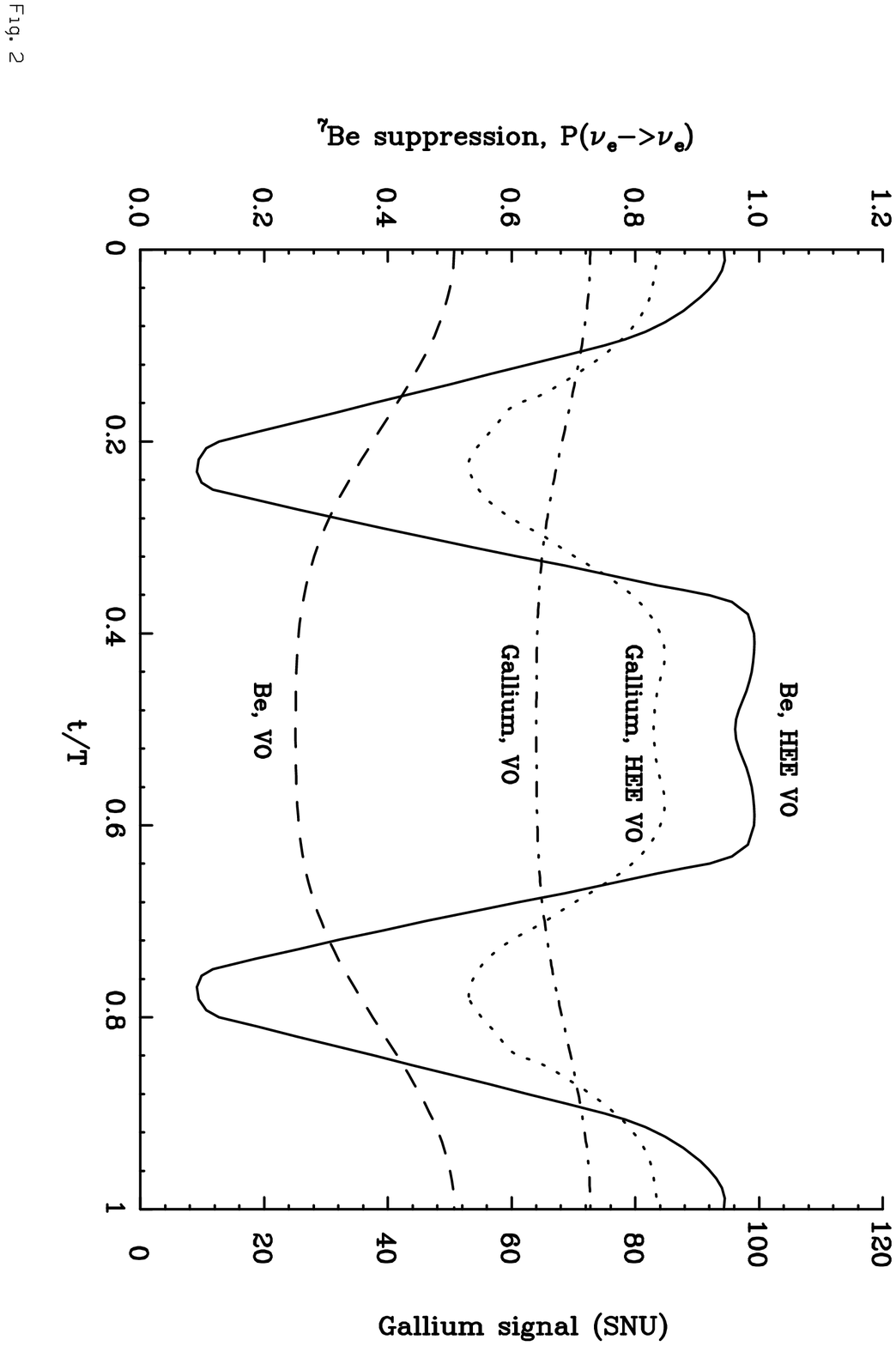,bbllx=36pt,bblly=36pt,bburx=576pt,bbury=756pt,%
height=22.5cm}
\end{figure}

\begin{figure}[c]
\psfig{figure=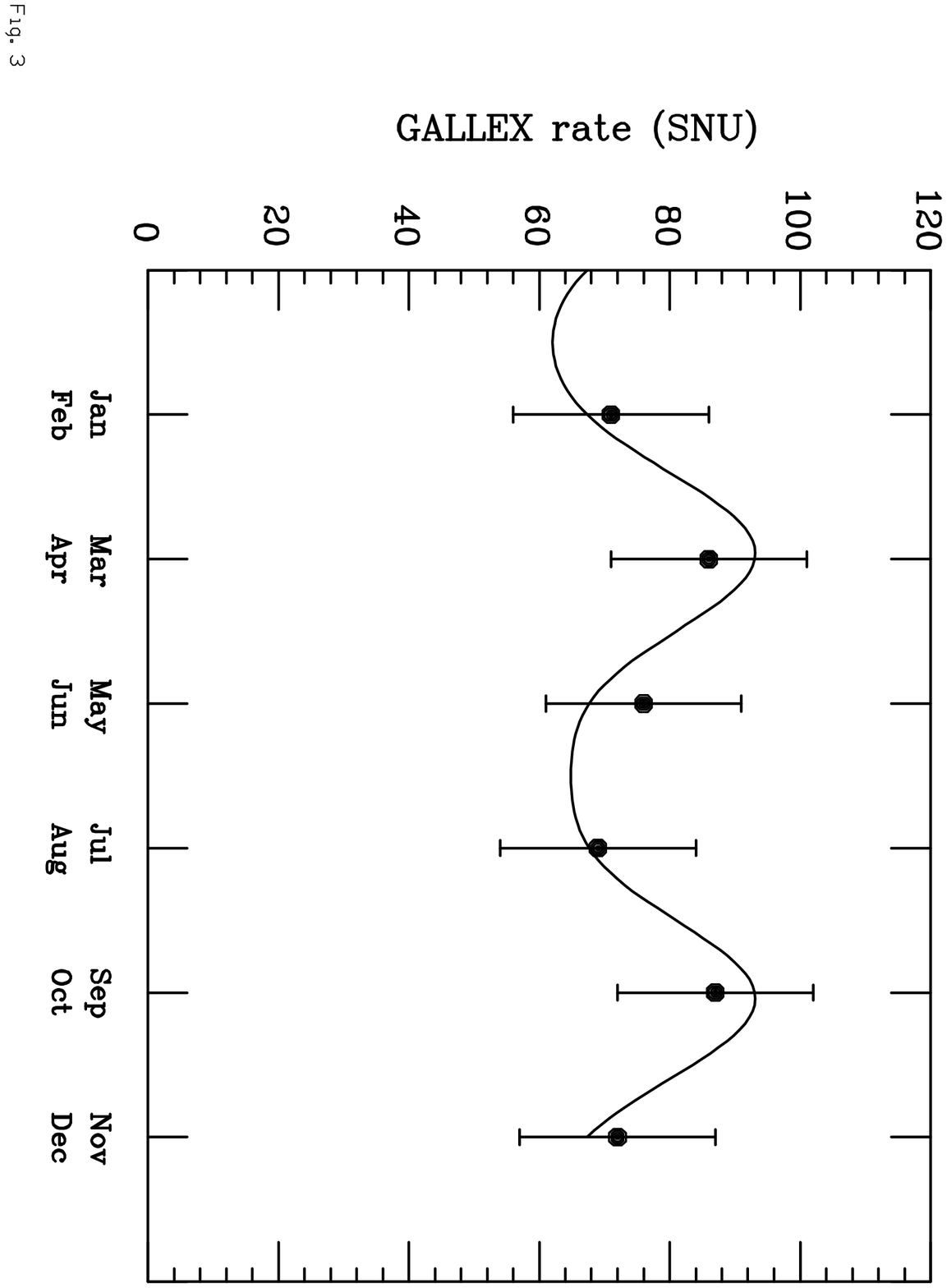,bbllx=36pt,bblly=36pt,bburx=576pt,bbury=756pt,%
height=22.5cm}
\end{figure}

\begin{figure}[c]
\psfig{figure=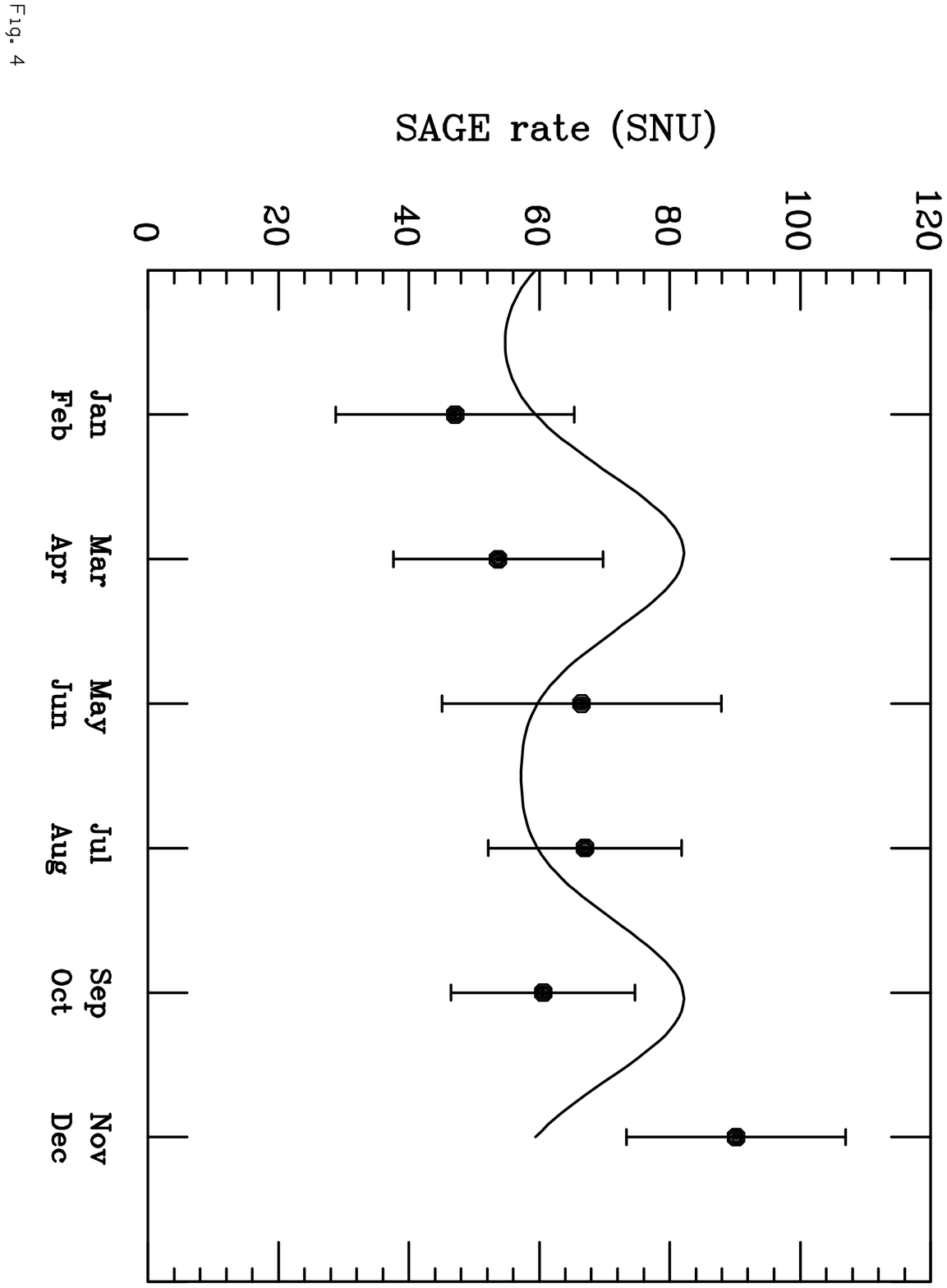,bbllx=36pt,bblly=36pt,bburx=576pt,bbury=756pt,%
height=22.5cm}
\end{figure}

\begin{figure}[c]
\psfig{figure=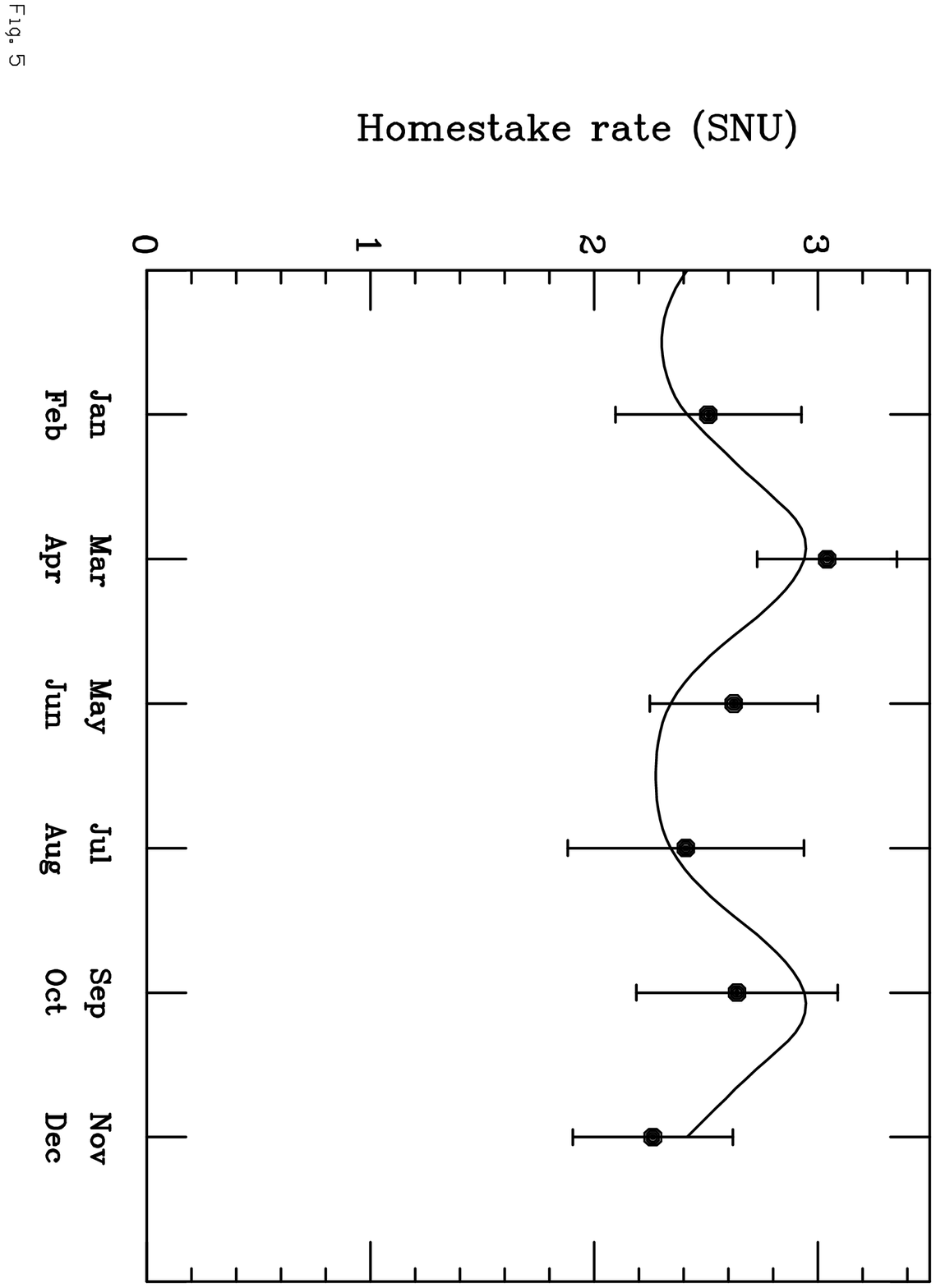,bbllx=36pt,bblly=36pt,bburx=576pt,bbury=756pt,%
height=22.5cm}
\end{figure}

\begin{figure}[c]
\psfig{figure=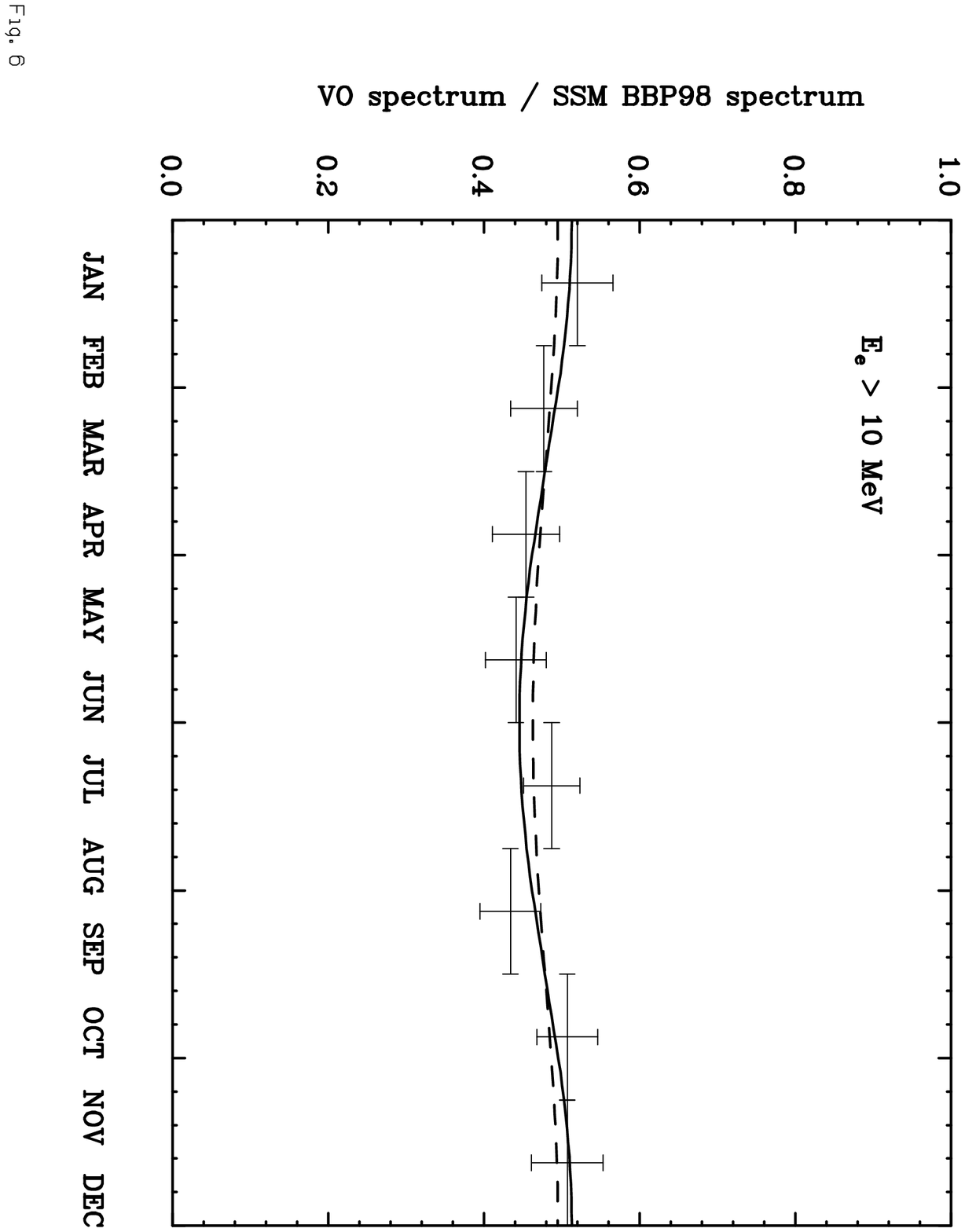,bbllx=36pt,bblly=36pt,bburx=576pt,bbury=756pt,%
height=22.5cm}
\end{figure}

\begin{figure}[c]
\psfig{figure=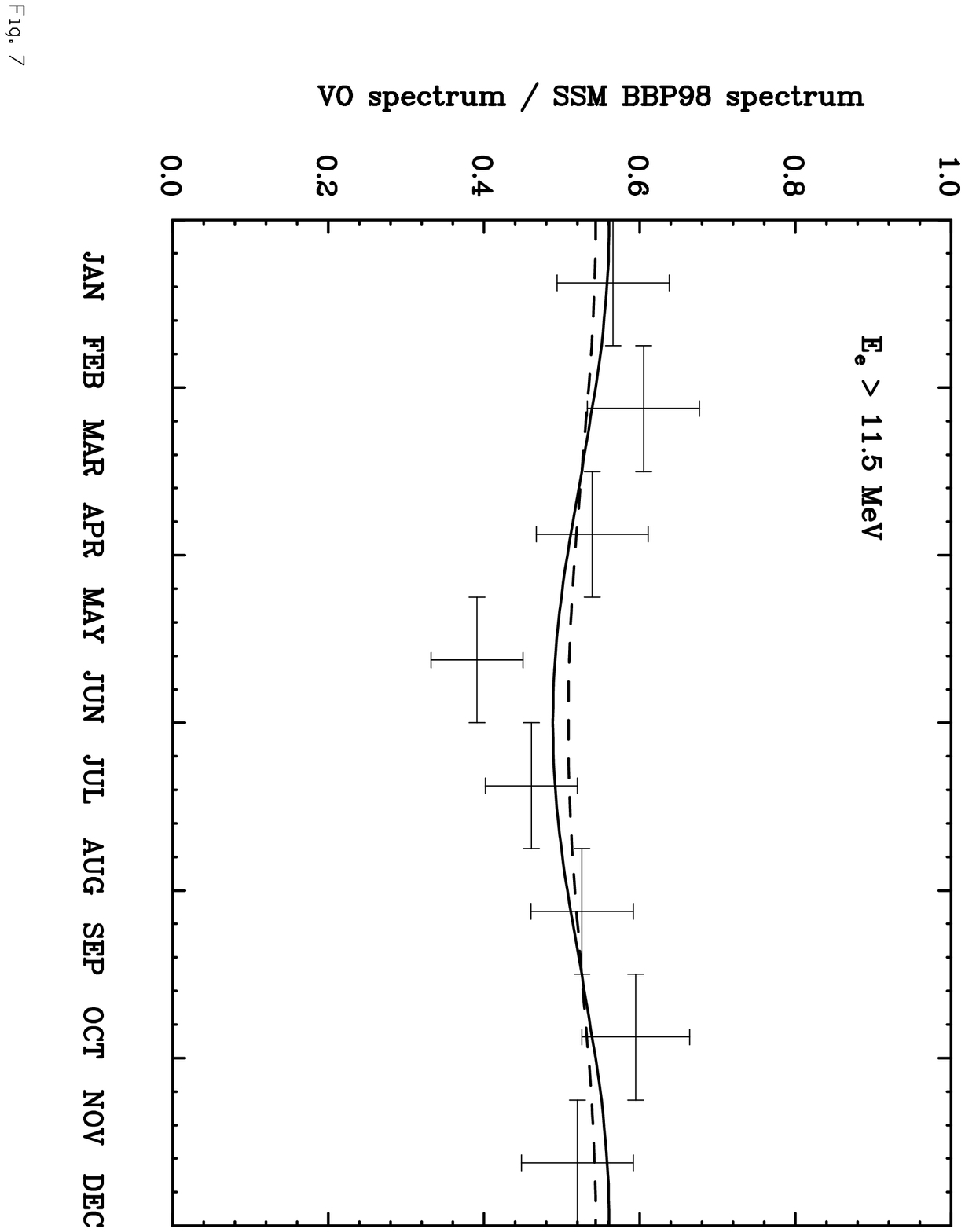,bbllx=36pt,bblly=36pt,bburx=576pt,bbury=756pt,%
height=22.5cm}
\end{figure}


\begin{thebibliography}{99}
\bibitem[*]{email1}
Electronic address: berezinsky@lngs.infn.it

\bibitem[\dagger]{email2}
Electronic address: fiorentini@fe.infn.it

\bibitem[\ddagger]{email3}
Electronic address: marcello.lissia@ca.infn.it

\bibitem{REF:SKAM}
SuperKamiokande Coll., Y. Suzuki, in : Neutrino 98, 
Proceedings of the XVIII International Conference on Neutrino
Physics and Astrophysics, Takayama, Japan, 4-9 June  1998,
Y. Suzuki and Y. Totsuka eds. To be published in 
Nucl. Phys. B (Proc. Suppl.).

\bibitem{REF:Totsuka}
Y.Totsuka, Talk at the 19th Texas Symposium on Relativistic Astrophysics,
Paris, 14 - 18 December 1998.

\bibitem{REF:Inou}
K.~Inoue , Talk at 8th Int. Workshop ``Neutrino Telescopes'', Venice , 23 -28 
February 1999.

\bibitem{REF:BKS}
J.~N.~Bahcall, P.~I.~Krastev, and A.~Yu.~Smirnov, Phys. Rev. {\bf D58},
                096016 (1998), hep-ph/9807216

\bibitem{REF:SNO}
A.~B.~McDonald, Nucl. Phys. B (Proc. Suppl.) {\bf 48}, 357 (1996).

\bibitem{escri}
R.~Escribano, J.~M.~Frere, A.~Gevaert, and D.~Monderen,
             Phys. Lett. {\bf B444}, 397 (1998), hep-ph/9805238.

\bibitem{REF:BaKr}
J.~N.~Bahcall and P.~I.~Krastev, Phys. Lett. {\bf B436}, 243 (1998),
         hep-ph/9807525.

\bibitem{REF:BDFR} 
G.~Fiorentini, V.~Berezinsky, S.~Degl'Innocenti, and B.~Ricci, 
        Phys. Lett. {\bf B444}, 387 (1998), astro-ph/9810083.

\bibitem{INT}
E.~G.~Adelberger, {\em at al.}, Rev. Mod. Phys. {\bf 70}, 1293 (1998).


\bibitem{REF:dvo}
V.~Berezinsky, G.~Fiorentini, and M.~Lissia, hep-ph/9811352 (1998).

\bibitem{REF:Bahc}
J.~N.~Bahcall,
{\em Neutrino Astrophysics} (Cambridge University Press, 1989);

\bibitem{REF:BiPe}
S.~M.~Bilenky and S.~T.~Petcov, Rev. Mod. Physics, {\bf 59}, 671 (1987).

\bibitem{REF:Turck}
S.~Turck-Chieze, W.~Daepen, E.~Fossat, J.~Provost, E.~Schatzman, 
and D.~Vignaud, Phys. Reports {\bf 230}, 57 (1993).

\bibitem{REF:Hax} 
W.~Haxton, Ann. Rev. Astron. Astroph. {\bf 33}, 459 (1995).

\bibitem{REF:HaLa95}
N.~Hata and P.~Langacker, Phys. Rev. {\bf D52}, 420 (1995).

\bibitem{REF:BK96}
J.~N.~Bahcall and P.~I.~Krastev, Phys. Rev. {\bf D53}, 4211 (1996).

\bibitem{REF:KP96}
P.~I.~Krastev and S.~T.~Petcov, Phys. Rev. {\bf D53}, 1665, (1996).

\bibitem{REF:HaLa97}
N.~Hata and P.~Langacker, Phys. Rev. {\bf D56}, 6107 (1997).

\bibitem{REF:Lisi}
B.~Faid, G.~L.~Fogli, E.~Lisi, and D.~Montanino, Astropart. Phys.
                      {\bf 10}, 93 (1999), hep-ph/9805293.

\bibitem{REF:KrSm}
P.~Krastev and A.~Smirnov, Phys. Lett. {\bf B338}, 282 (1994).

\bibitem{REF:BP98}
J.~N.~Bahcall, S.~Basu and M.~H.~Pinsonneault, Phys. Lett. {\bf B433},
            1 (1998), astro-ph/9805135.

\bibitem{REF:Pom}
I.~Ya.~Pomeranchuk (cited in~\cite{REF:BiPe}).

\bibitem{Valle}
P.~C.~de~Holanda, C.~Pena-Garay, M.~C.~Gonzales-Garcia, and J.~W.~F.~Valle,
hep-ph/9903473 (1999).

\bibitem{BKS1}
J.~N.~Bahcall, P.~I.~Krastev, and A.~Yu.~Smirnov, hep-ph/9905220 (1999).

\bibitem{REF:BiPo}
S.~M.~Bilenky and B.~M.~Pontecorvo, Phys. Rep. {\bf 41}, 225 (1978).

\bibitem{REF:GK}
S.~L.~Glashow and L.~M.~Kraus, Phys. Lett. {\bf B190}, 199 (1987).

\bibitem{REF:KrPe95}
P.~I.~Krastev and S.~T.~Petcov, Nucl. Phys. {\bf B449}, 605 (1995).

\bibitem{REF:GKK}
S.~L.~Glashow, P.~J.~Kerman, and L.~M.~Krauss, Phys. Lett. {\bf B445},
          412 (1998), hep-ph/9808470.

\bibitem{REF:MS}
S.~P.~Mikheev and A.~Yu.~Smirnov, Phys. Lett. {\bf B429}, 343 (1998).

\bibitem{REF:EHRLICH}
R.~Ehrlich, Phys. Rev. {\bf D18}, 2323 (1978).

\bibitem{REF:Rosen}
J.~M.~Gelb and S.~P.~Rosen, Phys. Rev. (Rapid Communication) {\bf D60},
011301 (1999), hep-ph/9809508.

\bibitem{REF:Hom}
B.~T.~Cleveland {\em et al.}, Ap.~J. {\bf 496}, 505 (1998). 

\bibitem{REF:Ki}
T.~Kirsten, Talk at  
the XVIII International Conference on Neutrino
Physics and Astrophysics, Takayama, Japan, 4-9 June  1998.

\bibitem{REF:Ga} SAGE collaboration, V.~N.~Gavrin {\em et al.}, Talk at  
the XVIII International Conference on Neutrino
Physics and Astrophysics, Takayama, Japan, 4-9 June  1998, to be published in 
Nucl. Phys. B (Proc. Suppl.).

\bibitem{REF:Suzuki}
Y.~Suzuki, Talk at 17th Int. Workshop on Weak Interactions and Neutrinos 
(WIN99),  Cape Town (South Africa) 24 - 30  January  1999.

\bibitem{REF:BOREXINO}
C.~Arpesella, {\em et al.}, BOREXINO proposal, eds.
G.~Bellini, R.~Raghavan {\em et al.} , Univ. of Milano, 1992.

\bibitem{REF:LENS}
R.~S.~Raghavan, Phys. Rev. Lett. {\bf 19}, 3618 (1997).

\bibitem{REF:Petcov}
M.~Maris and S.~T.~Petcov, hep-ph/9903303 (1999). 

\bibitem{BarWhi} 
V.~Barger and K.~Whisnant, hep-ph/9903262 (1999).

\end{thebibliography}
\end{document}